\documentclass{ieeeaccess}
\usepackage{cite}
\usepackage{amsmath,amssymb,amsfonts}
\usepackage{algorithmic}
\usepackage{graphicx}
\usepackage{textcomp}

\usepackage{microtype}
\usepackage{nameref}
\usepackage{hyperref}
\usepackage{babel}
\usepackage[utf8]{inputenc}

\NewSpotColorSpace{PANTONE}
\AddSpotColor{PANTONE} {PANTONE3015C} {PANTONE\SpotSpace 3015\SpotSpace C} {1 0.3 0 0.2}
\SetPageColorSpace{PANTONE}

\def\BibTeX{{\rm B\kern-.05em{\sc i\kern-.025em b}\kern-.08em
    T\kern-.1667em\lower.7ex\hbox{E}\kern-.125emX}}
\begin{document}
\history{Date of publication xxxx 00, 0000, date of current version xxxx 00, 0000.}
\doi{XXXXXXXXXXXX}

\title{Serious Games and AI: Challenges and Opportunities for Computational Social Science
}
\author{\uppercase{Jaime Pérez}
, \uppercase{Mario Castro
, and Gregorio López
} \IEEEmembership{Senior Member, IEEE}}
\address[]{Institute for Research in Technology, ICAI Engineering School, Universidad Pontificia Comillas, Madrid, 28015, Spain}
\tfootnote{This work has received funding from the European Union's Horizon 2020 research and innovation program under grant agreement No. 882828 (project RAYUELA). The sole responsibility for the content of this document lies with the authors and in no way reflects the views of the European Commission. This work has been partially supported by Grant PID2019-106339GB-I00 funded by MCIN/AEI/ 10.13039/501100011033.}

\markboth
{J. Pérez \headeretal: Preparation of Papers for IEEE TRANSACTIONS and JOURNALS}
{J. Pérez \headeretal: Preparation of Papers for IEEE TRANSACTIONS and JOURNALS}

\corresp{Corresponding author: Jaime Pérez (e-mail: jperezs@comillas.edu, ORCID: 0000-0001-6044-0022).}

\begin{abstract}
The video game industry plays an essential role in the entertainment sphere of our society. However, from Monopoly to Flight Simulators, serious games have also been appealing tools for learning a new language, conveying values, or training skills. Furthermore, the resurgence of Artificial Intelligence (AI) and data science in the last decade has created a unique opportunity since the amount of data collected through a game is immense, as is the amount of data needed to feed such AI algorithms. This paper aims to identify relevant research lines using Serious Games as a novel research tool, especially in Computational Social Sciences. To contextualize, we also conduct a (non-systematic) literature review of this field. We conclude that the synergy between games and data can foster the use of AI for good and open up new strategies to empower humanity and support social research with novel computational tools. We also discuss the challenges and new opportunities that arise from aspiring to such lofty goals.
\end{abstract}

\begin{keywords}
Serious Games, Artificial Intelligence, Computational Social Science, Novel Research Tools, Human behaviour
\end{keywords}

\titlepgskip=-15pt

\maketitle

\section{Introduction}
\label{intro}
\PARstart{G}{ames} have existed in all human societies and many other animal species. While some of the oldest board games, such as Go, Backgammon, or Checkers, are still played today, video games have become one of the most relevant forms of entertainment in our society, with budgets and profits far exceeding those of huge related industries such as cinema \cite{wijman2020global}. However, since the origin of games, they have had intentions and benefits beyond entertainment, such as teaching social norms, strengthening social bonds, or developing imagination and planning skills.

The rise of video games has had a remarkable social impact, transforming mentalities and helping to establish new patterns of social interaction \cite{rogers2016video}. A prominent example of this trend is the gamification that our lives have experienced \cite{Koivisto2019rise}, from the workplace (e.g., \textit{Habitica}, \textit{LifeUp}) to romantic relationships (e.g., \textit{Tinder}, \textit{Grindr}) or education (e.g., \textit{Kahoot!}, \textit{Duolingo}). One of the main reasons games have such a tremendous impact on players is due to interactivity, an almost unique feature over other cultural or artistic elements. This attribute encourages higher motivation, engagement, and empathy levels than in other media. It is noteworthy that at the same time as the video game industry is rising, the board game industry continues to grow as well \cite{arizton2020board}. We can draw a clear conclusion from all these facts: our society is highly gamified; we love to play games, and they have enormous potential to transform how we see the world in a much more profound way than we are often aware of.

\begin{figure*}[hbt!]
  \center
  \caption{Graphical overview of the paper. The blue boxes are the applications found for serious games and AI applied to them. The red box indicates the challenges faced by this union for its use as a research tool. And the green box indicates promising lines of work in this direction.}
  \includegraphics[width=0.93\textwidth,keepaspectratio]{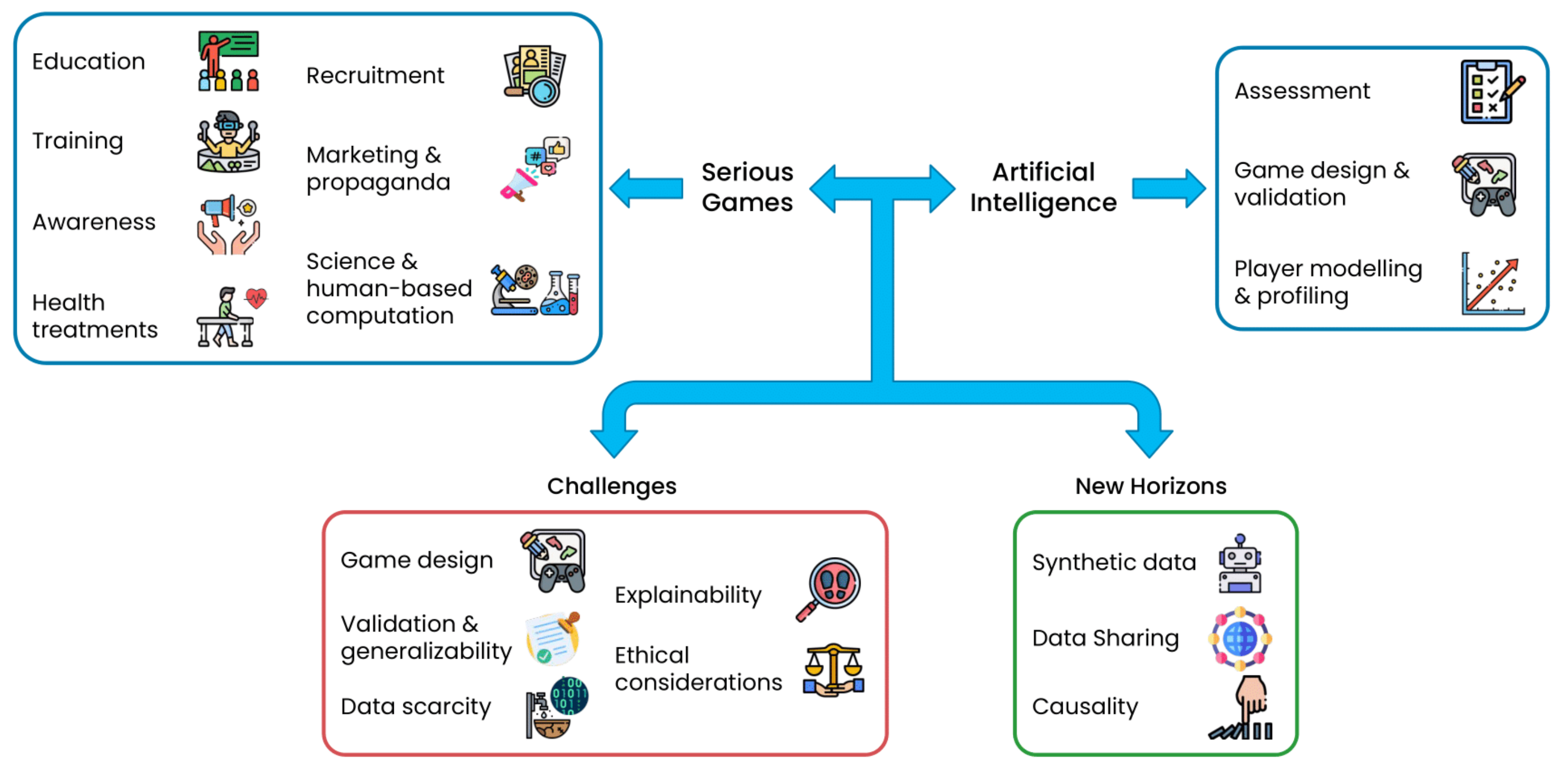}
  \label{fig:abstract}
\end{figure*}

Some games ---referred to as \emph{serious}~\cite{Abt1987serious}--- are explicitly designed for a primary purpose beyond pure entertainment (e.g., learning new skills, conveying values, awareness-raising). However, being entertaining is part of their attractiveness. The first serious games were released in a wide range of formats, from sports to board games (e.g., \textit{Monopoly}, \textit{Suffragetto}), so this concept precedes
the digital era. 

The current re-emergence of serious games has coincided with the eruption of Artificial Intelligence (AI), one of the most impressive game-changers in the history of humanity. Nowadays, and increasingly so, almost every entertainment element and digital product are at the service of data analysis and AI algorithms, and games are no exception. Especially given that the amount of data available via video games far exceeds any other artistic element. AI is already transforming society through large digital platforms, social networks, and recommender systems. The most widespread use of these tools is in marketing, meticulously analyzing our patterns and tastes to sell us products and capture our attention as much as possible. 

AI has demonstrated its potential to analyze and improve our understanding of the dynamics of our societies, social interactions, as well as individual and collective behaviors. For this reason, we firmly believe that the synergy between serious games and AI offers an exceptional window of opportunity for large-scale, non-invasive, and inexpensive social studies, leveraging their disinhibition and entertainment effects, along with interactivity, to collect large amounts of meaningful data. Moreover, games' "casual" and playful nature can help break down conventional communication boundaries, encouraging participants to interact openly and discuss topics that might otherwise be complicated or too sensitive. In Figure \ref{fig:abstract} we can find a visual overview of this paper's content to guide and facilitate the reading of the document. Including applications of serious games (Section \ref{sec:app}), the role of AI in them (Section \ref{sec:AI}), and the new lines of work and challenges opening up for employing them as research tools (Section \ref{sec:challenge}), especially in the computational social sciences \cite{ComputationalSS}. Finally, Section \ref{sec:conclusion} presents the main conclusions drawn from this research.

\section{Applications of Serious Games}
\label{sec:app}
The upsurge that serious games have been experiencing in recent years may lead us to think this is a new phenomenon. However, the origin of serious games dates back to the 1970s. Clark C. Abt is credited for coining the term \emph{serious games}, defining them as "\textit{games with an explicit and carefully thought-out educational purpose that are not intended to be played primarily for amusement}". Clark C. Abt studied the potential of games as a vehicle for political, educational, or marketing ideas. Another of the leading figures in the history of serious games is Ian Bogost, the author of seminal books on the theory behind them, such as "\textit{Persuasive Games: The expressive power of video games}" \cite{bogost2010persuasive}. His research has leveraged video games for political, educational, and business use in the 21st century. 

Even though both concepts mirror the same social phenomenon, it is relevant to highlight the distinction between gamification and serious games. Gamification consists of using and integrating game elements into non-game concepts, while serious games refer to the design of entire games for non-playful primary purposes. Although both are concepts from the last century, they have resurfaced in the academic and commercial arenas in recent years.

Among the first serious video games, we find examples where they are employed to convey particular values (e.g., \textit{Captain Bible in the Dome of Darkness}, \textit{The Oregon Trail}, \textit{Mobility}), disease awareness (e.g., \textit{Captain Novolin}), or military training (e.g., \textit{Bradley Trainer}). Nevertheless, the line between ''normal'' and ''serious'' games is quite blurred regarding serious games used to convey specific beliefs or ideologies. Like any artistic or intellectual creation, video games always carry an implicit political and philosophical perspective. For example, popular video games such as \textit{Papers Please} or \textit{This War of Mine} convey strong political messages that raise fundamental questions. Yet, they were not developed under the idea of being explicit ''serious games''. On the other hand, video games such as \textit{The Sims} or \textit{SimCity} are very politically charged. Still, they are not usually perceived as such since they represent a situation closer to our day-to-day life.

Focusing on serious games that consider themselves as such and have been designed for that purpose, we find many fields where they have demonstrated their usefulness on numerous occasions:

\subsection*{Education}
In this section, we focus on serious games designed for the player to learn a series of concepts of a specific subject. To do so, the players must demonstrate their knowledge during the game and score their performances. Education has been one of the main focuses of action for serious games, based on the principle that learning while having fun is possible and efficient. This field has been explored so extensively that success and failure factors have even been analyzed in depth \cite{zhonggen_meta_2019} \cite{ravyse_success_2017}. Prominent examples of building STEM skills might include \textit{Garfield's Count Me In} \cite{Garfield2021}, \textit{Minecraft: Education Edition} \cite{Minecraft2018}, the \textit{Kahoot DragonBox} maths apps \cite{DragonBox} and the \textit{LightBot} coding apps \cite{Lightbot}. Serious games for educational purposes have also become popular in higher medical education \cite{sharifzadeh2020health}, although some authors question their usefulness at such high educational levels, possibly serving as complements to more traditional learning methods \cite{gorbanev2018systematic} \cite{haoran2019serious}.

\subsection*{Training}
Closely related to education, this category refers to games designed for players to learn and practice specific skills that will enable them to perform those actions in the real world with improved safety, confidence, and knowledge. This approach is widely used in companies where human failure is critical or costly. One of the best-known examples is flight simulators, such as \textit{Microsoft Flight Simulator} \cite{2020_microsoft}, where aspiring pilots must spend hours practicing before flying an actual commercial aircraft. There are also notable examples of training healthcare professionals \cite{wang_systematic_2016}, cybersecurity trainees \cite{hendrix2016game} \cite{TiohCyber2017}, and law enforcement agencies or military forces \cite{akhgar2019serious} \cite{samvcovic2018serious}. Another widespread use is training to manage complex business situations or the administration of teams and resources, used both in actual private companies \cite{gamelearn2015} \cite{larson_serious_2020} and universities \cite{HarvardSimulators2021} \cite{Zoomsim}.

\subsection*{Awareness}
Thanks to the almost unique characteristic of interactivity, games evoke deep levels of empathy, making them an ideal vehicle to convey an awareness of relevant social issues. A classic example is \textit{Darfur is Dying} \cite{Darfur2006}, which sought to tell the story of the humanitarian crisis in the Darfur region of South Sudan. However, we can find examples on a wide range of topics, such as drug consumption and trafficking \cite{stapinski2018pure} \cite{code2018fight}, cyberbullying \cite{CALVOMORATA2020}, gender equality \cite{barrera2020review}, misinformation \cite{educsci9020155} \cite{Harmony2021}, climate change \cite{flood2018adaptive}, and environmental sustainability \cite{den2018evaluating} \cite{aubert2018review} \cite{stanitsas2019facilitating} \cite{johnson2017gamification}.

\subsection*{Health Treatments}
This category is framed in healthcare but focuses more on patients than professionals. Well-known examples might be the \textit{Wii Fit} and \textit{Brain Training} games, which aim to have fun and stay fit (physically and mentally) simultaneously. Other notable examples can be found in the field of mental health therapy \cite{Ferguson2021} \cite{fleming2017serious}, increasing self-efficacy and physical activity in people with chronic diseases \cite{Holly2020} \cite{bossen2020effectiveness}, helping the learning process and support of children with autism \cite{fridenson2017emotiplay} \cite{khowaja2019serious}, palliative care and memory training for the elderly and/or people with dementia \cite{Huansheng2020Dementia} \cite{nguyen2017impact}, and guidance and motivation in rehabilitation processes \cite{lopes2018games} \cite{KARAMIANS2020885} \cite{AYED2019103909} \cite{MEIJER20181890}. Notably, in 2020 the US Food and Drug Administration approved the first video game-based treatment, \textit{EndeavorRx}, targeting children between the ages of eight and twelve with certain types of  Attention Deficit Hyperactivity Disorder (ADHD) \cite{CNN2020game}.

\subsection*{Recruitment}
\label{recruitment}
If we combine games' interactivity with players' ability to make decisions in a well-designed environment, we can infer some behaviors or aspects of the players' abilities with reasonable confidence. For this reason, serious games have also been used to optimize the recruitment process in private companies \cite{buzady2019fligby} \cite{loreal2014} and even in military forces \cite{america2002army}. In these games, players are presented with complex situations where they must make decisions and act under certain constraints or pressures. A recent notable example is the \textit{CodinGame}\footnote{CodinGame \url{https://www.codingame.com}} platform, where users practice their programming skills while playing, and many tech companies recruit profiles they find interesting. Another great example is the \textit{GT Academy}\footnote{GT Academy \url{https://www.gran-turismo.com/es/academy/}}, a competition in which the best players of a car racing video game have the opportunity to become professional drivers.

\subsection*{Marketing \& Propaganda}
\label{marketing}
When the game is developed primarily for marketing purposes, it is often known as an ''advergame''. This category of games aims to convey ideas and create desires unintrusively and easily customizable way. It should not be confused with games that introduce advertising during gameplay for economic profit. The principal medium for these advergames is smartphones due to their proliferation, ease of development, and everyday use among young people. 

Major brands such as \textit{Volkswagen}, \textit{Magnum}, \textit{Chupa Chups} or \textit{M\&Ms} have developed advergames. Concerning the \nameref{recruitment} category, in some cases, companies seek to present and profile themselves through these games to attract new employees and trainees or discover talent. 

Likewise, there have also been attempts to use video games as a tool to disseminate electoral campaigns, such as the video game \textit{Corbyn Run} \cite{Corbyn2019}, or to encourage citizen participation in public decisions \cite{hazMadrid2017} \cite{schouten2016playful}.  

\subsection*{Science \& Human-based Computation}
This category encompasses games to advance scientific knowledge in some way. One of the most common approaches is employing human players to perform seemingly trivial tasks, either too costly, too complex, or unfeasible with finite computational resources. These tasks may include labelling data, transcribing text, using common sense, or activities based on the human experience. 

One of the first examples of this category was ''\textit{The ESP Game}'' \cite{von2004labeling}, in which players, grouped in pairs, had to guess the photo labels their partner had come up with. Google's reCAPTCHA\footnote{reCAPTCHA \url{https://www.google.com/recaptcha/about/}} is a recent example that has followed this approach of using human players to label images while identifying legitimate users for accessing online resources. Another successful example was ''\textit{EteRNA}'' \cite{eterna2010}, where players had to design RNA sequences that fold into a particular form. The solutions were evaluated to improve computer-based RNA folding prediction models. Other prominent examples might be ''\textit{Foldit}'' \cite{foldit2008} to predict protein structures, ''\textit{Eyewire}'' \cite{eyewire2012} to map retinal neurons, ''\textit{MalariaSpot}'' \cite{malariaspot2016} to help diagnose malaria cases, ''\textit{Phylo}'' \cite{kawrykow_phylo_2012} to optimize alignments of nucleotide sequences, or ''\textit{Quantum Moves}'' \cite{quantum2012} to improve how atoms move in a quantum computer.

\section{Role of AI and Data Science in Serious Games}
\label{sec:AI}

Games have long been the test bed for AI as they provide a controlled environment with simple rules for algorithms to learn sophisticated strategies. However, in recent years data science and AI have found a different use for games as sources of vast amounts of player data. Moreover, thus be able to get relevant information about the human players that may be useful both outside and inside the game itself.

Nevertheless, serious games are a particular branch of the gaming industry, so the AI and analysis techniques and their purposes differ noticeably. The significant heterogeneity in the goals of serious games also implies significant technical differences among them. Despite this heterogeneity, we can discern main branches encompassing all major applications of AI and analytics in serious games:

\subsection*{Assessment}
\label{assessment}

Game-based assessment is a fruitful field in serious games \cite{KimAssessment2019}, primarily used in education, training, and recruitment. Players are scored based on their knowledge or skills in a particular subject. Pellegrino \textit{et al.} \cite{Pellegrino2001knowing} stipulate three primary purposes of assessment: (i) to assist learning (formative assessment), (ii) to evaluate the player's capabilities, and (iii) to evaluate programs. In general, collecting, analyzing, and extracting information through educational serious games is known as Game Learning Analytics \cite{FreireGLA2016}.

The main difference with traditional evaluation methods or test gamification is that game-based assessment also uses in-game and interaction data (e.g., response times) to evaluate the player. Numerous authors have demonstrated the utility of using additional in-game data to evaluate students \cite{liu_learning_2017} \cite{kiili_evaluating_2018} \cite{AlonsoLessons2019} or to predict learning results \cite{AlonsoPredicting2020} \cite{hernandez_lara_applying_2019}. It has also been successfully tested in recruitment processes \cite{LandersAssessment2022}. Although nowadays, they are more of a complement to the traditional exam-based assessment.

The techniques used are very diverse, from simple descriptive statistics and correlations to supervized machine learning algorithms (e.g., linear regression, decision trees, Naive Bayes, Neural Networks) \cite{AlonsoApplications2019} \cite{gomez2022systematic}. More rarely, some papers use knowledge inference with Bayesian networks \cite{RoyBayes2019} \cite{shute2016assessing}, which explicitly allows the application of psychological or mental state models, but a flawed model will negatively influence the results significantly.

This branch of AI applications in serious games is one of the most researched and developed, thanks to the technology push changing how education is delivered. However, much work still needs to be done, especially in demonstrating that they can be better than traditional approaches \cite{PamelaAssessmentJungle2017}.

\subsection*{Game Design \& Validation}
Game design is planning the content, rules, and mechanics of a game to create valuable interactive experiences. The large number of artistic and technical factors involved in this process make any analytical information about the players extremely valuable. Game validation employs data and evidence to verify and calibrate the game tasks and their difficulty. In the case of serious games, in addition to maintaining engagement, we also want to ensure that the game meets its primary objective (e.g., to train players in a particular skill, increase awareness of an issue, etc.).

Data-driven serious game design has flourished in academia in recent years, where we can find successful examples of the use of analytical techniques to design, improve, personalize, and validate these games \cite{Hicks_Puzzle2016} \cite{calvo_validation_2020} \cite{AlonsoLessons2019} \cite{Freire2016} \cite{Peddycord2017}. This category is also closely related to the previous one (\nameref{assessment}), as it is almost essential to use data-driven validation during the game development stage to calibrate the players' evaluation \cite{tong2016serious} \cite{kowalewski_validation_2017}. Such analytics can go a step further to adapt in real-time the difficulty of the game \cite{hendrix2018implementing} \cite{hocine2015adaptation} and even detect when the player is frustrated \cite{defalco2018detecting}.

In this category, due to the particular aspects of design and validation of each game, the most commonly used techniques are descriptive statistics and visualizations \cite{kang_using_2017} \cite{alario_datadriven_2020} \cite{Cano_Using_2018}, Randomized Control Trials (to test the usefulness of the intervention) \cite{calvo_creating_2021} \cite{calvo_validation_2020} and unsupervized machine learning algorithms (to find similar types of players and common patterns in the game) \cite{calvo_validation_2020}.

Using these analytical techniques enables creators and researchers to ensure that their games are entertaining, engaging, and well-designed to fulfill their objectives.

\subsection*{Player modeling \& Profiling}
Player modeling is the creation of computational models to detect, predict and characterize the human player attributes that manifest while playing a game \cite{yannakakis2018artificial}. These models can be any mathematical representation, rule set, or probability set that maps parameters to observable variables and are built on dynamic information obtained during game-player interaction. On the other hand, player profiling usually refers to categorizing players based on static information that does not alter during gameplay (e.g., personality, cultural background, gender, age). Despite their dissimilarities, these two concepts can complement each other, contributing to more reliable player models. 

The main objective of studying players is to understand their cognitive, affective, and behavioral patterns. Recent advances in AI have demonstrated an impressive ability for these same goals that player modeling sets out to achieve. Although, at the moment, there is a significant lack of interpretability in complex models. AI for player modeling is a perfectly good fit solely when explainability is not a hard constraint.

Hooshyar \textit{et al.} \cite{SurveyModeling2018} conducted a systematic literature review that profoundly analyzes the computational and data-driven techniques used for player modeling between the period of 2008 to 2016. As this is such a broad and promising field, the variety of algorithms used is immense: descriptive statistics and correlations, path/network analysis, supervized learning (e.g., Neural Networks, Linear Regression, Hidden Markov Models, Decision Trees), unsupervized learning (e.g., k-means, Linear Discriminant Analysis, Self-Organizing Map), probabilistic algorithms (e.g., Bayesian / Markov Networks), evolutionary methods (e.g., Genetic algorithms), reinforcement learning methods (e.g., Multi-armed bandits), etc. Most of the computational methods used are model-free, meaning they do not impose strict assumptions on the model. However, there are also some model-based approaches (e.g., Bayesian hierarchical models) \cite{Streicher_Bayes2022} \cite{Kim_Computational2018} that yield more interpretable and explicit models (e.g., psychological or cognitive) than those which do not impose strict assumptions on the model. For instance, these models can infer the player's hidden parameters or mental states.

Player modeling can be helpful both inside and outside the game itself. The most straightforward goal is to improve the game design, tailoring the content to increase engagement and enhance the gaming or learning experience \cite{DELIMA201832}. Although outside of serious games, we find some prominent examples, such as \textit{Left 4 Dead} \cite{valve2008left}, where an AI tracks player behavior and adapts future waves of enemies to maintain rhythm and tension. Perhaps the most famous example is the video game \textit{Silent Hill Shattered Memories} \cite{climax2009silent}, which uses a psychological approach where an AI system tries to manipulate players' emotions using the \textit{Five Factor Model} of personality \cite{digman1990personality}. Outside the game itself, the most common use of player modeling in the gaming industry is for personalized marketing campaigns, since the commercial sector is very interested in understanding customer behaviors and preferences. In these cases, the games are often presented as free to play in exchange for an intrusion into personal privacy \cite{DREIER2017328}. Besides the "advergames" discussed in the section \nameref{marketing}, a famous example outside serious games is \textit{Farmville} \cite{willson2015zynga}, which monitored the players' behavior to adapt \textit{Amazon} marketing campaigns to them. This business model is particularly hazardous for younger users, its main target.

In academia, especially in psychology, experiments have been conducted using games (serious and non-serious) for research, but primarily focusing on analyzing how the player's personality is projected in the gameplay patterns \cite{GamesPersonality2011} \cite{Yee_Introverted_2011} \cite{Halim_Profilin_2019} \cite{denden_implicit_2018} \cite{MCCORD201995}. However, studying psychological characteristics or phenomenology using serious games seems an up-and-coming field, especially if we introduce AI techniques into the equation.

\section{Challenges and New Horizons}
\label{sec:challenge}

In the previous sections, we have discussed the main applications of serious games and the current trends in their synergies with data science and AI. In this section, we take up the argument outlined in the introduction about the great potential of serious games together with AI to serve as research tools, particularly in computational social sciences \cite{ComputationalSS}, examining the most critical challenges and promising new lines of work to meet this objective.

As argued in the \nameref{intro} section, games allow research to be entertaining, provide high levels of empathy, and have a disinhibition effect that is highly sought after in social investigations. Games can evoke dynamic and complex emotions in players, the manifestations of which are difficult to capture with the traditional approaches of empirical psychology, affective computing, or cognitive modeling research. This is primarily due to their ability to introduce the player to a continuous mode of interaction, which could generate complex cognitive, emotional, and behavioral reactions \cite{yannakakis2018artificial}. Therefore, the use of serious games as research tools may contribute to the advancement of human-computer interaction and the progress of our knowledge of human experiences.

We can already find some splendid examples of the use of games as large-scale social research tools, such as \textit{The Moral Machine Experiment} \cite{awad_moral_2018}, which uses a gamified environment to explore the moral dilemmas surrounding autonomous cars. To do so, they use the framework of the classic trolley problem and study participants' responses to variations in different parameters (e.g., number of people who would die, age, gender, etc.) and the cross-cultural differences in this decision-making \cite{AwadMoralVariations2019}. We can also find some noteworthy examples that use serious games to explore collaborative and trusting behaviors \cite{pereda_group_2019} \cite{PoncelaDyadic2016}, understand preferences for charity donations\footnote{MyGoodness! \url{https://www.my-goodness.net/}}, or even fight cybercrime \cite{RAYUELA}.

On the other hand, the latest advances in AI allow us to analyze vast amounts of data and find patterns or behaviors that would be very difficult to observe with traditional analytical methods. So far, the main application given to large AI models that study our interactions through social networks and personal data is for marketing purposes and generating monetary value \cite{Ma_Marketing_2020}. This practice has been done almost since the beginning of social networks, without considering the negative social consequences it could have, particularly for children and adolescents \cite{Keles_SocialMedia_2020} \cite{Abi_Smartphones_2020}. With this paper, we also aim to contribute humbly to the ''AI for Good''\footnote{AI for Good Global Summit \url{https://aiforgood.itu.int/}}\textsuperscript{,}\footnote{AI for Good \url{https://ai4good.org/}} movements. We are at a critical social, cultural, and economic moment. We must start to consider the uses of AI that can benefit society and each individual in particular. We firmly believe that IA has the potential to help us live better and also to know ourselves better. Furthermore, to achieve great goals that improve our society, it is essential to unite forces between different branches of science (e.g., sociology, psychology, engineering, computer science, AI, etc.), and we believe that games represent such an excellent vehicle for this purpose.

However, in order for serious games to be able to meet these major goals, they must face some critical \textbf{challenges}:

\begin{itemize}
    \item \textit{Game design}: Whether a game can serve as a valuable research tool depends strongly on whether it has good design and playability. Designing a game is a complex process involving many artistic and technical aspects that can not be wholly rationalized from a scientific standpoint.
    \item \textit{Validation and generalizability}: One of the most complicated aspects of using serious games as a means of research is demonstrating that their results are as valid as traditional methods. Although we already have numerous examples in some branches, such as game-based assessment or the reflection of player morality into in-game moral dilemmas \cite{Weaver_Mirrored_morality} \cite{Sven_influence_moral}, there is still a long way to go in this aspect. This is also because each game (and its purpose) is different from the others and therefore requires individual validation in most cases.
    \item \textit{Data scarcity}: In recent years, it has become clear that to take full advantage of AI, we need large amounts of data to feed it. Apart from a few exceptional cases \cite{awad_moral_2018}, serious games academic experiments suffer from small, biased, and heterogeneous datasets. If we aspire to use them as social research tools, we must find ways to get more participants, make the best use of available data or establish appropriate methods of sharing sensitive data.
    \item \textit{Explainability}: Many of today's AI tools can be highly complex, if not completely opaque (so-called black box models). The general trend in computer science also drives this to focus more on prediction than explanation. However, aspiring to use these tools to study human and social behavior implies a deep understanding of the outcomes that AI provides. While considerable progress has been made in explainable AI techniques, many hurdles still exist \cite{bruijn_perils_2022}.
    \item \textit{Ethical considerations}: When dealing with personal data (whether anonymized or not) and AI, we must seek unequivocal ethical standards. The potential benefits must outweigh the risks, as the participants' safety and well-being must be the top priority, especially when dealing with data from children or people at risk of exclusion. Achieving these standards is genuinely complex because computer scientists and social scientists tend to have different approaches to research ethics \cite{salganik2019bit}. 
\end{itemize}

Despite the challenges mentioned above, we can also find promising \textbf{new horizons} and future lines of work regarding the interplay between serious games and AI:

\begin{itemize}
    \item \textit{Synthetic data}: The AI field has extensive experience in developing agents that aim to win a game \cite{Hasselt_DQN_2016}. However, in recent years, we are also experiencing the emergence of novel synthetic data generation techniques capable of modeling or mimicking human behavior in some aspects \cite{Hussein_Imitation_2017}, and impressive new data augmentation techniques such as Generative Adversarial Networks \cite{goodfellow_GAN}. Concerning the challenge of data scarcity, this is a promising line of work in which we could make the most of the limited data available and build models that help us better understand players' motivations in decision-making.   
    \item \textit{Data sharing}: The field of computational social science has faced many difficulties in finding and sharing open data, especially from private companies \cite{Lazer_CSS_obstacles_2020}. However, the field of serious games is in a much more advantageous position in this respect, as it does not involve such an amount of sensitive data. Moreover, using anonymization and privacy-preserving algorithms has proven to be very useful in recent years. With this promising line of work, we can address the poor sample size that traditional social science has had and share meaningful data from serious games at scale to enhance collaboration and motivate research.
    \item \textit{Causality}: The social sciences have traditionally prioritized interpretable explanations of human behavior, mainly invoking causality through randomized controlled trials. However, as powerful as these techniques are, they are also very costly in terms of resources and money. On the other hand, computer scientists have traditionally been more concerned with developing accurate predictive models, whether or not they correspond to causal/interpretable mechanisms. Nevertheless, in the last years, we are experiencing a resurgence of computational causality techniques \cite{mcelreath2020statistical} \cite{glymour2016causal}, even from observational data (i.e., quasi-experiments) \cite{liu_quantifying_2021}, allowing us to explain with greater robustness the workings of the systems under study. Besides, it makes explicit the assumptions of the computational model and the scientist performing it, helping us to make research more open to discussion and to rethink plausible alternatives for existing explanations. If our ultimate goal is to better understand individual and collective human behavior, it is critical to integrate predictive and explanatory approaches to scientific research \cite{hofman_integrating_2021}.
\end{itemize}

\section{Conclusions}
\label{sec:conclusion}

Gaming, both for entertainment and utility purposes, has been indispensable throughout the development of humankind. The flourishing of AI in recent times, coupled with the vast amounts of meaningful data that can be collected and transmitted through games, creates a unique window of opportunity to use serious games as tools for social research.

In this paper, we have reviewed serious games' main applications and their synergies with AI. We can already find numerous successful examples of serious games in education, science, business and social interests. The great potential of games to transform society should not be underestimated and deserve more and deeper inquiry. In addition, we have identified some challenges and promising new lines of work for using serious games as research tools. By doing so, we aim to motivate researchers to pursue these lines of work and help them to identify potential applications of serious games for beneficial social objectives. We also want to encourage interdisciplinary research, which is essential in this field of science, and which we firmly believe is how the future of science should ultimately be.

We are at a critical juncture as a society, where we are beginning to realize that we need to change the motivations and goals by which we make progress. AI is a game changer that can bring immense benefits or harm to society. It is time to start breaking new ground in using these technologies for the common good. What better way to do it than by playing?

\bibliographystyle{ieeetr}
\bibliography{access.bbl}

\begin{IEEEbiography}[{\includegraphics[width=1in,height=1.25in,clip,keepaspectratio]{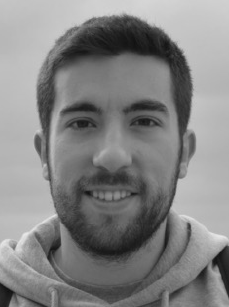}}]{Jaime P\'erez} received the B.S. and the M.Sc. degree in Telecommunication Engineering from the Universidad Politécnica de Madrid (UPM) in 2018 and 2021, respectively. \\ Currently, he is pursuing a PhD in Engineering Systems Modeling at the Institute for Research in Technology, Universidad Pontificia Comillas, within the framework of the EU H2020 project RAYUELA  (empoweRing and educAting YoUng pEople for the Internet by pLAying). His areas of interest are Serious Games, Machine Learning, Deep Learning, Synthetic Data, and Applied Artificial Intelligence.
\end{IEEEbiography}

\begin{IEEEbiography}[{\includegraphics[width=1in,height=1.25in,clip,keepaspectratio]{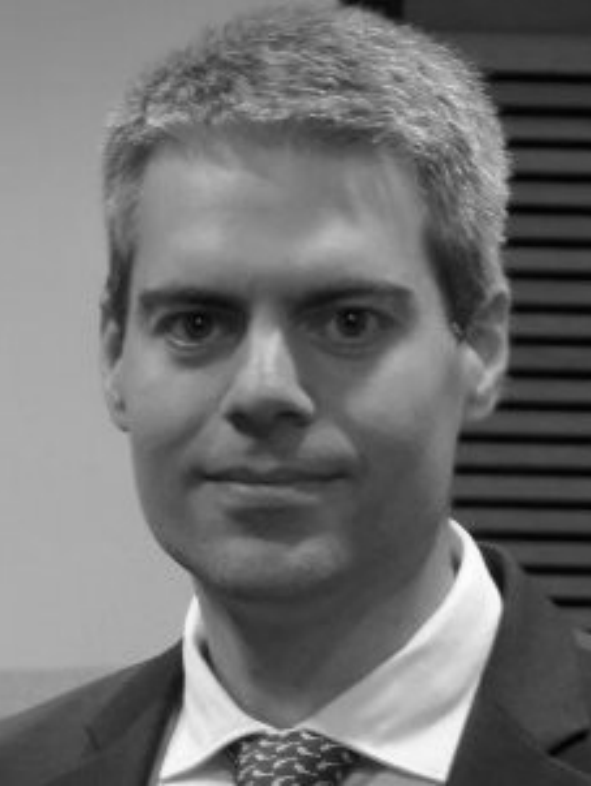}}]{Mario Castro} holds a Ph.D. in Physics from the Universidad Complutense de Madrid (2001). He has been the Principal Investigator of 5 Spanish Ministry of Science research projects. He has participated in 10 projects, including 3 European projects within the framework of Marie Sklodowska-Curie Actions (of which he has been the IP of the Comillas node) and H2020.\\
These projects' central theme was applying Statistical Mechanics methods to model complex systems, with particular emphasis on modeling and predicting experimental data. His primary research approach translates analytical and computational methods into actual experiments. In this sense, most of his research has been carried out in collaboration with experimental groups from all over the world, from Immunology to Computational Social Science. He has been a visiting researcher at the Los Alamos National Laboratory (LANL) and completed a sabbatical stay at the University of Leeds (2016-2017), of which he is currently Visiting Professor.
\end{IEEEbiography}

\begin{IEEEbiography}[{\includegraphics[width=1in,height=1.25in,clip,keepaspectratio]{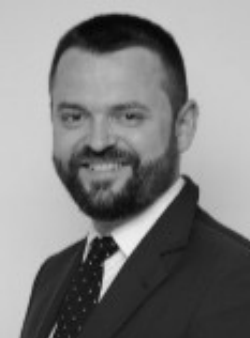}}]{Gregorio L\'opez} (SM'20) received his PhD in Telecommunications Engineering from Universidad Carlos III de Madrid (UC3M) in 2014. He currently works as Assistant Professor in the ICAI Engineering School of Comillas Pontifical University, where he also serves as the coordinator of the Cybersecurity MSc, and as Senior Researcher in the Institute for Research in Technology. He gathers wide experience in close-to-market research gained through his participation in more than 10 national and European research projects. As a result of his research activity, he holds an European Patent and has published more than 20 papers in top-tier conferences and journals. His current research interests revolve around the optimization of Machine-to-Machine (M2M) communications networks based on analysis and simulation, cybersecurity, and data analytics for the Internet of Things (IoT), and the use of technology and the Internet, being currently the coordinator of the European H2020 project RAYUELA (empoweRing and educAting YoUng pEople for the Internet by pLAying), which addresses this latter topic.  
\end{IEEEbiography}

\EOD

\end{document}